\documentclass[amsmath,amssymb,floatfix,pra,reprint]{revtex4-1}
\usepackage{graphicx}
\usepackage{physics}
\usepackage{color}
\usepackage{amsmath}
\usepackage{fullpage}
\usepackage{hyperref}
\usepackage{float}
\usepackage{times}
\usepackage{soul}

\hypersetup{
    colorlinks = true,
    allcolors = [rgb]{0,0,0},
}

\newcommand{\tp}{\tau_{\mathrm{p}}}
\newcommand{\ts}{\tau_{\mathrm{s}}}
\newcommand{\fe}{f_{\mathrm{e}}}
\newcommand{\eff}{\eta}
\newcommand{\noise}{\nu}
\newcommand{\nsr}{\mu_1}
\newcommand{\taub}{\bar{\tau}}
\newcommand{\OD}{\mathrm{OD}}
\newcommand{\ODs}{\mathrm{OD}_\mathrm{stat.}}

\begin{document}

\title{Fast, noise-free memory for photon synchronization at room temperature}

\author{Ran Finkelstein}
\thanks{These authors contributed equally to this work}
\affiliation{Department of Physics of Complex Systems, Weizmann Institute of Science,Rehovot 7610001, Israel}
\author{Eilon Poem}
\thanks{These authors contributed equally to this work}
\affiliation{Department of Physics of Complex Systems, Weizmann Institute of Science,Rehovot 7610001, Israel}
\author{Ohad Michel}
\affiliation{Department of Physics of Complex Systems, Weizmann Institute of Science,Rehovot 7610001, Israel}
\author{Ohr Lahad}
\affiliation{Department of Physics of Complex Systems, Weizmann Institute of Science,Rehovot 7610001, Israel}
\author{Ofer Firstenberg}
\email{To whom correspondence should be addressed; E-mail:  ofer.firstenberg@weizmann.ac.il.}
\affiliation {Department of Physics of Complex Systems, Weizmann Institute of Science,Rehovot 7610001, Israel}

\begin{abstract}

Future quantum photonic networks require coherent optical memories for synchronizing quantum sources and gates of probabilistic nature. We demonstrate a fast ladder memory (FLAME) mapping the optical field onto the superposition between electronic orbitals of rubidium vapor. Employing a ladder level-system of orbital transitions with nearly degenerate frequencies simultaneously enables high bandwidth, low noise, and long memory lifetime. We store and retrieve 1.7-ns-long pulses, containing 0.5 photons on average, and observe short-time \emph{external} efficiency of 25\%, memory lifetime ($\mathbf{1/e}$) of 86 ns, and below $\mathbf{10^{-4}}$ 
added noise photons. { Consequently, coupling this memory to a probabilistic source would enhance the \emph{on-demand} photon generation probability by a factor of 12,} the highest number yet reported for a noise-free, room-temperature memory.
This paves the way towards the controlled production of large quantum states of light from probabilistic photon sources.
\end{abstract}
\maketitle

Large quantum states of light, where many photons occupy multiple modes in a coherent superposition of different configurations, are the backbone of photonic quantum communication, metrology, and computation \cite{OpticalQuantumReview}. Optical photons are easily transmitted over complex networks and do not suffer from thermal noise at ambient temperature, making photonic quantum information processing an appealing paradigm \cite{LOQC}. However, the interaction between photons in common optical materials is extremely weak, leading to single-photon sources and two-photon gates that must rely on measurement and are therefore probabilistic. This renders the scaling-up of quantum photonic networks an exponentially hard problem.

Two approaches to this challenge are currently pursued. One approach focuses on developing new materials and systems for \emph{deterministic} operation, such as single emitters strongly coupled to photonic structures \cite{Pan16,Barak14,Rempe16,Lukin16} or cold ensembles of cooperative emitters \cite{Ofer13}. 
The second approach, which we follow here, focuses on actively synchronizing \emph{probabilistic} elements in a repeat-until-success strategy \cite{LOQC,Nunn13}. Here, every operation is repeated until it heralds a success, and its output is stored; when all operations are successful, their outputs are synchronously retrieved. This strategy requires memory modules that can efficiently store quantum states of light and retrieve them on demand, without additional noise \cite{QOMReview10,QOMReview15}.
Indeed, many types of quantum-optical memories have been demonstrated, motivated by the need for long term storage in quantum repeaters \cite{QOMReview10,opticalQCommReview} as well as for synchronizing high-bandwidth sources \cite{Nunn13,QOMReview15}. None of these demonstrations, however, meets the combined requirements of high external efficiency, long lifetime, and low added noise.
\begin{figure*}[tb]
\centering
\includegraphics[width=1.0\textwidth]{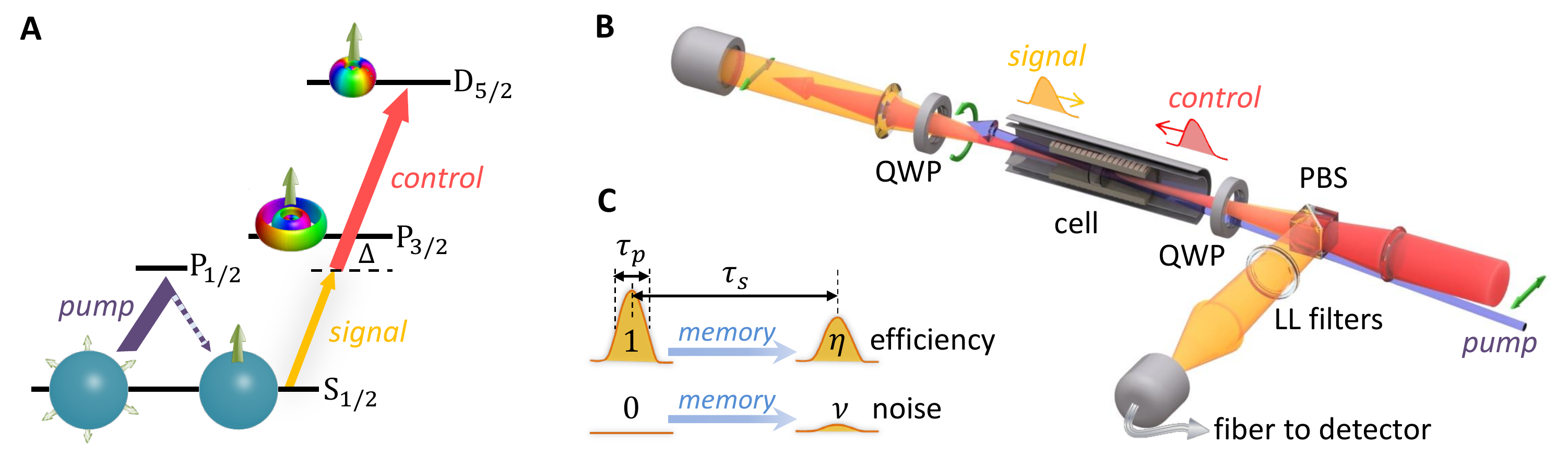}
\caption{\label{fig1} FLAME scheme. \textbf{(A)} A ladder level structure comprising purely orbital transitions (the surface colors display the phase structure of the orbitals $5s$, $5p$, and $5d$) is achieved by optical pumping (purple) of the nuclear and electronic spins (green arrows) to the maximally-polarized state. Nonzero detuning $\Delta$ from the intermediate level can be introduced. \textbf{(B)} To keep the ladder within the maximally-polarized subspace, the counter-propagating \textit{signal} and \textit{control} beams are circularly-polarized using quarter-wave plates (QWP; polarizations shown by green arrows). The collimated optical \textit{pump} beams enter the cell at a small angular deviation. After storage and retrieval, a polarizing beam-splitter (PBS) picks out the signal. Scattered control light and spontaneous emission are filtered-out spectrally by laser-line (LL) filters and spatially by a single-mode fiber coupled to the photodetector. \textbf{(C)} The parameters governing the synchronization capability of the memory are pulse duration $\tp$, memory lifetime $\ts$, retrieval efficiency $\eff$, and noise $\nu$.}
\end{figure*}

Here we demonstrate a fast ladder memory (FLAME) satisfying these requirements at ambient temperatures.
FLAME utilizes a ladder level scheme of electronic orbitals, in our case the $5s$, $5p$, and $5d$ orbitals of warm rubidium atoms, as illustrated in Fig.~\ref{fig1}(a). A strong \emph{control} pulse  induces the coherent absorption of the \emph{signal} pulse in the medium, thereby mapping the signal field onto the spatial field of quantum coherence between the lower and upper orbitals. A subsequent control pulse retrieves the signal via stimulated emission. 

Very recently, Kaczmarek \emph{et al.}~\cite{Kris17} introduced the off-resonant cascaded absorption (ORCA) protocol, which is a far-detuned FLAME. They used cesium vapor to store and retrieve single photons from a heralded down-conversion source and demonstrated the preservation of non-classical statistics. However, the large energy mismatch of the ladder levels in cesium limited the memory lifetime to a few nanoseconds. Here we implement FLAME in rubidium, obtaining a memory lifetime much longer than in cesium while maintaining high bandwidth and low noise. We show that FLAME works both on resonance ($\Delta=0$) and off resonance ($\Delta=1.15$ GHz) with the intermediate ladder level. Despite the differences between these two regimes \cite{Raman_mem_th_Nunn07,Giacobino2012,NovikovaPRL2007}, we obtain high external efficiency in both.

The experimental arrangement is depicted in Fig.~\ref{fig1}B.
We prepare the ensemble by polarizing the atoms using circularly polarized ($\sigma^+$) optical pumping beams at 795 nm. To keep the polarized atoms within the maximally-polarized subspace of the 5S$_{1/2}-$5P$_{3/2}-$5D$_{5/2}$ ladder, we use $\sigma^+$ signal at 780 nm and $\sigma^+$ control at 776 nm (Fig.~\ref{fig_scheme}). Transitions within this subspace are purely orbital, even when $\Delta$ and the control Rabi frequency $\Omega$ are not much larger than the hyperfine splitting.

Three main ingredients combine to explain the physics behind the success of FLAME. First, the fast time variation of the control field breaks the delay-bandwidth constraint of time-invariant linear resonant systems \cite{LipsonNatPhys07,BoydScience2017}, thus enabling coherent absorption of signal pulses substantially shorter than the atomic resonance lifetime \cite{Raman_mem_th_Nunn07,Giacobino2012}. Second, with entirely orbital transitions and no need for spin flips, the FLAME operation is essentially independent of hyperfine and fine interactions~\cite{Eilon15}, making its bandwidth fundamentally limited only by the distance to adjacent orbitals.
Third, the large $\Omega$ (and optionally large $\Delta$) diminish the effect of Doppler broadening and by that efficiently engage in the process all velocity groups of the thermal ensemble. This makes the resonant optical depth in our experiments ($\OD=13$) ideally as efficient as if the atoms were stationary 
(thus equivalent to $\ODs\approx 800$) and comparable to that of large ultra-cold ensembles \cite{YuEIT13,BuchlerGem16}. The optical depth and control intensity contribute equally to the storage efficiency for a given pulse duration \cite{Raman_mem_th_Nunn07}, making their experimental availability the only practical limitation on the FLAME bandwidth.

Together with high bandwidth, a main appeal of FLAME over $\Lambda$-type (ground-state) memories is its immunity to four-wave-mixing noise; in $\Lambda$-type configurations, the control may spontaneously Raman-scatter to form spurious spin-waves, which are then retrieved as noise photons \cite{Michelberger15}. This process is absent in a ladder configuration \cite{Kris17}. Furthermore, spurious excitations of the 5P and 5D levels are heavily suppressed by the frequency mismatch between the two ladder transitions and by the negligible excitation probability of the optical transition at room temperature.

In addition to reducing spurious excitations, the wavelength difference between the signal and control fields allows for their separation by commercial interference filters. This enables high setup transmission, so that the memory (external) efficiency approaches the (internal) efficiency of the storage process itself. However, the wavelength difference inevitably leads to a spatially varying phase of the stored coherence and thus to dephasing due to ballistic atomic motion. This dephasing, known as \emph{residual} Doppler broadening, scales with the wave-vector difference between the signal and control and is thus minimized in the counter-propagation geometry. In rubidium FLAME, the wave-vector difference is $\sim 0.5\%$ (coherence wavelength of $\sim 150$ $\mu$m), providing an excellent trade-off between the above competing factors. The resulting expected dephasing time is about 130 ns at the cell temperature of $100^\circ$C, comparable to the 240 ns radiative lifetime of the 5D level. While this lifetime is shorter than that of most ground-state memories \cite{KuzmichEIT05, EIT_mem1, EIT_mem2, Michelberger15,YuEIT13,BuchlerGem16}, it is much larger than the inverse bandwidth of FLAME, as required for efficient synchronization.

The memory parameters governing the synchronization performance are illustrated in Fig.~\ref{fig1}C. For any storage time $t$, the \emph{external} efficiency $\eff_t$ is the ratio between the number of retrieved photons arriving at the detector and that of the incoming photons. The memory lifetime $\ts$ (defined by $\eff_{t=\ts}=\eff_0e^{-1}$) and pulse duration $\tp$ determine the fractional delay $\ts/\tp$.  The \emph{effective} fractional delay \mbox{$\fe=\eff_0 \ts/\tp$} quantifies the synchronization capacity { for an array of source-memory units \cite{Nunn13}. This is the enhancement factor, with respect to a bare source, of the probability that a single unit of such an array would generate a photon at the readout time.}    
%
%
%

\begin{figure}[tb]

\centering
\includegraphics[width=\columnwidth]{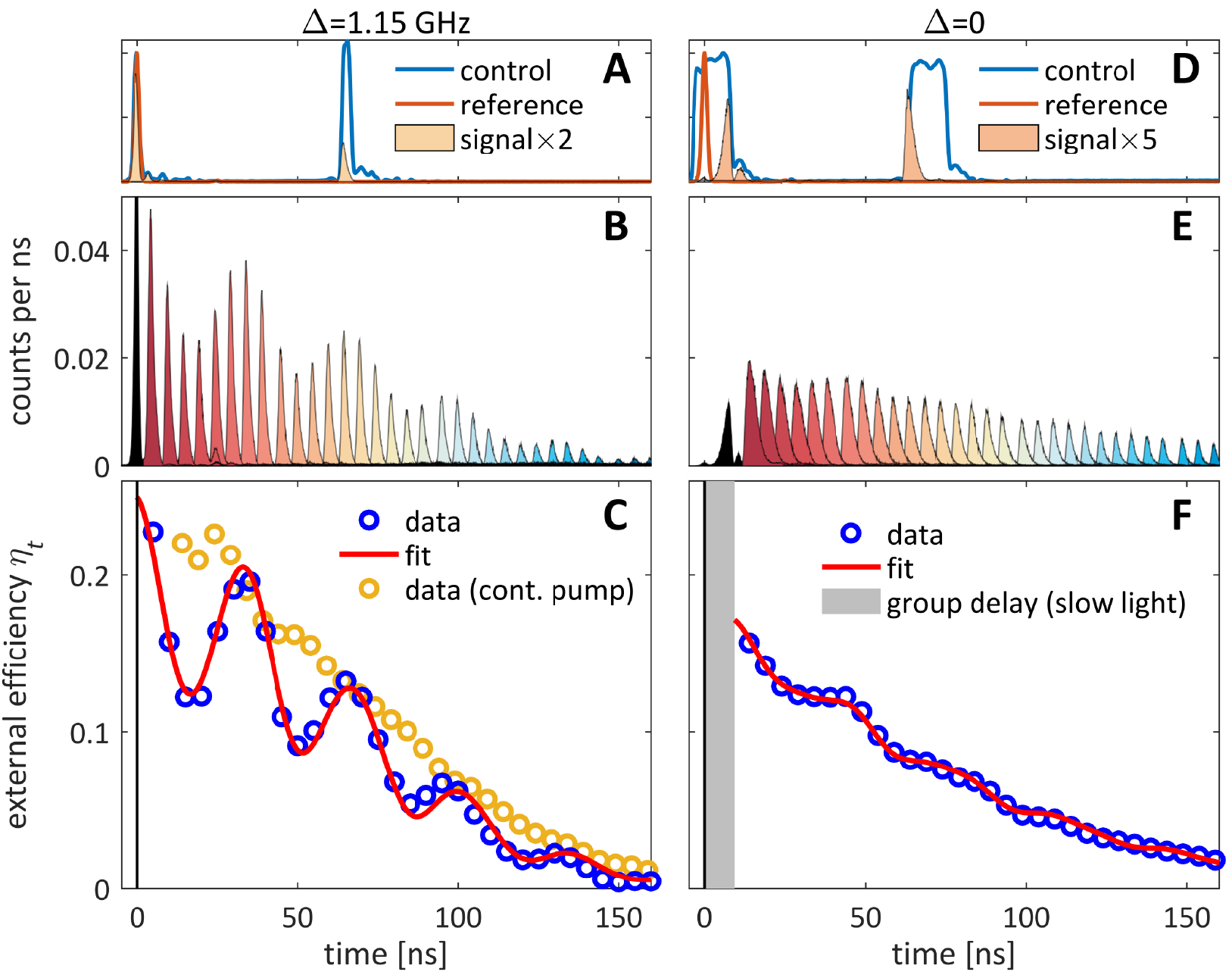}
\caption{\label{fig2} FLAME operation for different storage times $t$, off resonance (left) and on resonance (right). \textbf{(A,D)} Typical pulse sequence, presented for $t=40$~ns. The incoming pulse is shown for reference. \textbf{(B,E)} Traces from the single-photon counter for different storage times (colors). The blackened areas mark the portion of the leaked signal. \textbf{(C,F)} Decay of memory efficiency with $t$. Continuous optical pumping (yellow symbols in \textbf{C}) demonstrates the vanishing of beating for a fully polarized ensemble (while introducing more noise). Gray area in \textbf{(F)} marks the delay of the signal due to reduced group velocity while the control pulse is on. 
}
\end{figure}

Finally, the noise $\noise$ is the number of photons arriving at the detector absent an incoming signal. The noise-to-signal ratio \mbox{$\nsr\equiv\noise/\eff_0$} quantifies the contamination of the retrieved state for one input photon \cite{Saunders16}.
The characterization of the memory lifetime and efficiency in the off-resonance case $\Delta = 1.15$ GHz is shown in Fig.~\ref{fig2}A-C. We store laser pulses with a duration of $\tp=1.7$~ns, containing 0.5 photons on average. These measurements are repeated on resonance ($\Delta=0$) with $\tp=1.85$ ns, as shown in Fig.~\ref{fig2}D-F. The pulse durations, and thus the demonstrated bandwidth of $\sim250$~MHz, are limited by our driving electronics ({see Materials and Methods}). We adjust the timing and duration of the control pulses (see Fig.~\ref{fig2}A,D) for maximal storage efficiency. For the on-resonance case, we arrive at a long square control pulse, within which the signal exhibits a substantial group delay ($\approx9$~ns), typical for storage via electromagnetically induced transparency (EIT) \cite{EIT_mem1}. Additionally, the pulses retrieved on resonance are wider by $\sim 60\%$ than the input pulse, compared to just $\sim10\%$ widening for off resonance. 

We turn off the optical pumping $1.2~\mu$s prior to the storage in order to reduce the fluorescence from excited atoms (see Supplementary Material Sec. S2). During the off time, unpolarized atoms enter the interaction region, reducing the average polarization and enabling the excitation of additional hyperfine sublevels in 5D$_{5/2}$. The beats in Fig.~\ref{fig2}C (also visible in Fig.~\ref{fig2}F) are due to interference between the retrieval amplitudes from these sublevels. Indeed, the beating visibility vanishes for continuous pumping (Fig.~\ref{fig2}C, yellow) and conversely increases without optical pumping  (Sec. S2). 
 
We fit the measured efficiency to a simple model accounting for the beats (with the known hyperfine frequencies) and for inhomogeneous and homogeneous decays ({ see Materials and Methods}). The dominant decay source is found to be motional dephasing. For $\Delta=1.15$ GHz, we extract from the fit the short-time efficiency $\eff_0=0.25(1)$ and lifetime $\ts=86(2)$ ns. This yields an effective fractional delay $\fe=\eff_0 \ts/\tp= 12.6$, the largest yet reported for noise-free memories at ambient temperature. For $\Delta=0$, we extract $\eff_0=0.171(4)$ and $\ts=82(1)$ ns, yielding $\fe=7.6$. 
\begin{figure}[tb]
\centering
\includegraphics[width=\columnwidth]{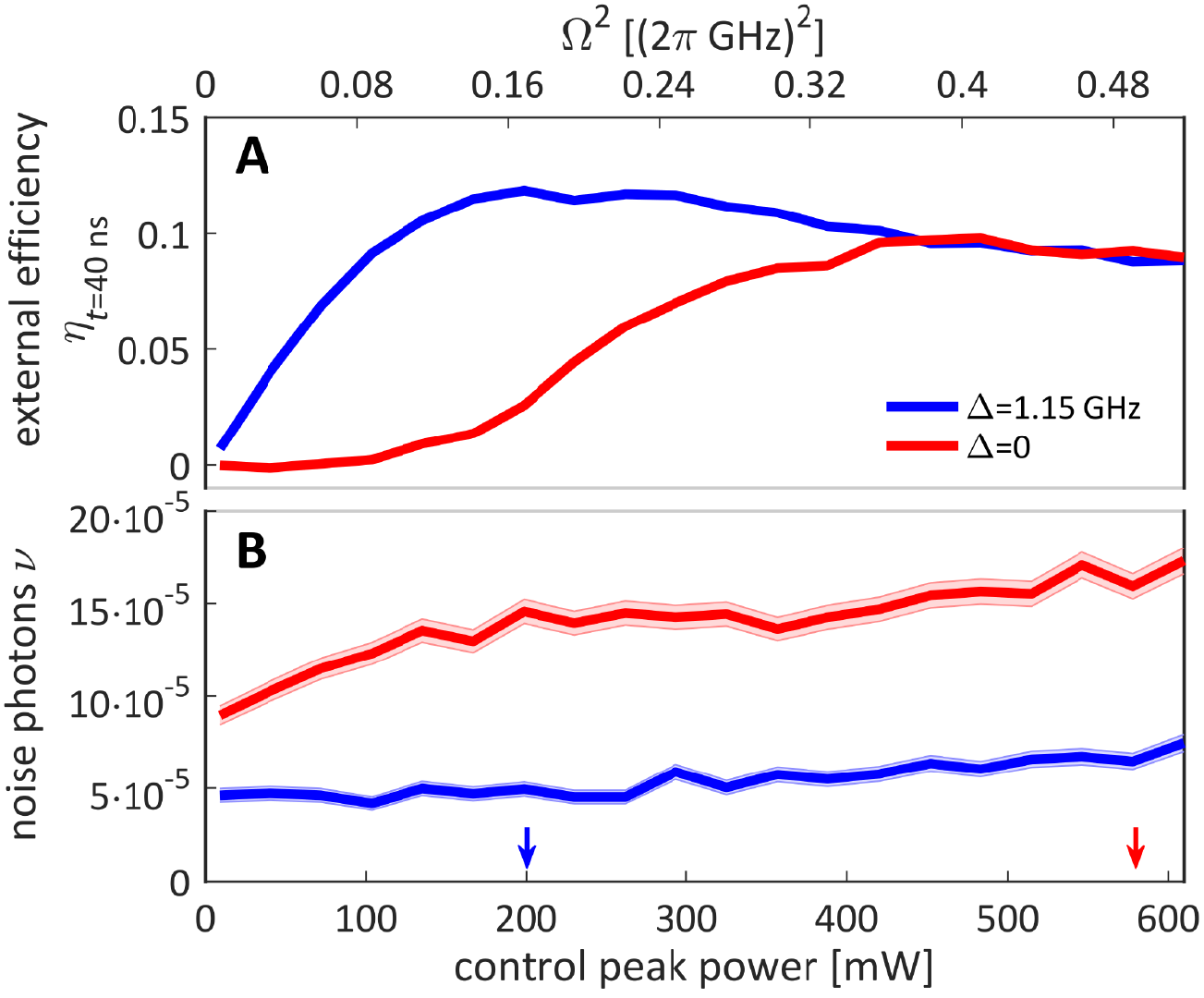}
\caption{\label{fig3}Efficiency and noise dependence on control power (bottom axis) or control Rabi frequency $\Omega$ (top axis). \textbf{(A)} Memory efficiency for $t=40$~ns. \textbf{(B)} Total photons in the collection window absent an incoming signal (shaded areas are $1\sigma$ statistical uncertainty). Note that the collection window is larger by $45\%$ in the $\Delta=0$ case (for accommodating the wider retrieved signal). The arrows mark the operating powers of Fig.~\ref{fig2}.}
\end{figure}

The noise $\noise$ is measured by repeating the experiment with no input signal. We observe no dependence of the noise on storage time and find $\nsr=\noise/\eff_0=11(1)\cdot 10^{-4}$ for $\Delta=0$ and $\nsr=2.3(3)\cdot 10^{-4}$ for $\Delta=1.15$ GHz, an order of magnitude lower than current ground-state memories.

The efficiency and noise as a function of control power are shown in Fig.~\ref{fig3} for a fixed storage time. The $\Delta=0$ case reaches the maximal efficiency at higher control power. The noise $\noise$ scales linearly with control power, affirming the absence of four-wave mixing, which would have yielded a quadratic dependence. We attribute the noise to unfiltered control photons and residual fluorescence due to the optical pumping (Sec.~\ref{SM_noise}).
These measurements, in combination with the direct verification in Ref.~\cite{Kris17} that the memory preserves anti-bunching, establish the suitability of FLAME for quantum synchronization applications.

The differences between on and off resonant FLAME, evident in the optimal duration (Fig.~\ref{fig2}A,D) and power (Fig.~\ref{fig3}A) of the control pulses, imply that the storage dynamics in the two regimes are somewhat distinct. It appears that off-resonant FLAME, or ORCA\cite{Kris17}, performs slightly better. On the other hand, they exhibit comparable efficiencies at high control power, once the resonant absorption has been overcome in the $\Delta=0$ case. 
Group-velocity dispersion moderately widens the pulse on resonance, but is expected to vanish at higher control power \cite{NovikovaPRL2007}; the off-resonance storage allows for higher degree of control over the retrieved pulse shape \cite{Fisher16}, and the observed minor widening can be circumvented by fine-tuning the shape of the control pulses.
Future work could explore the optimal regime of operation for FLAME.

\begin{figure}[tb]
\centering
\includegraphics[width=\columnwidth]{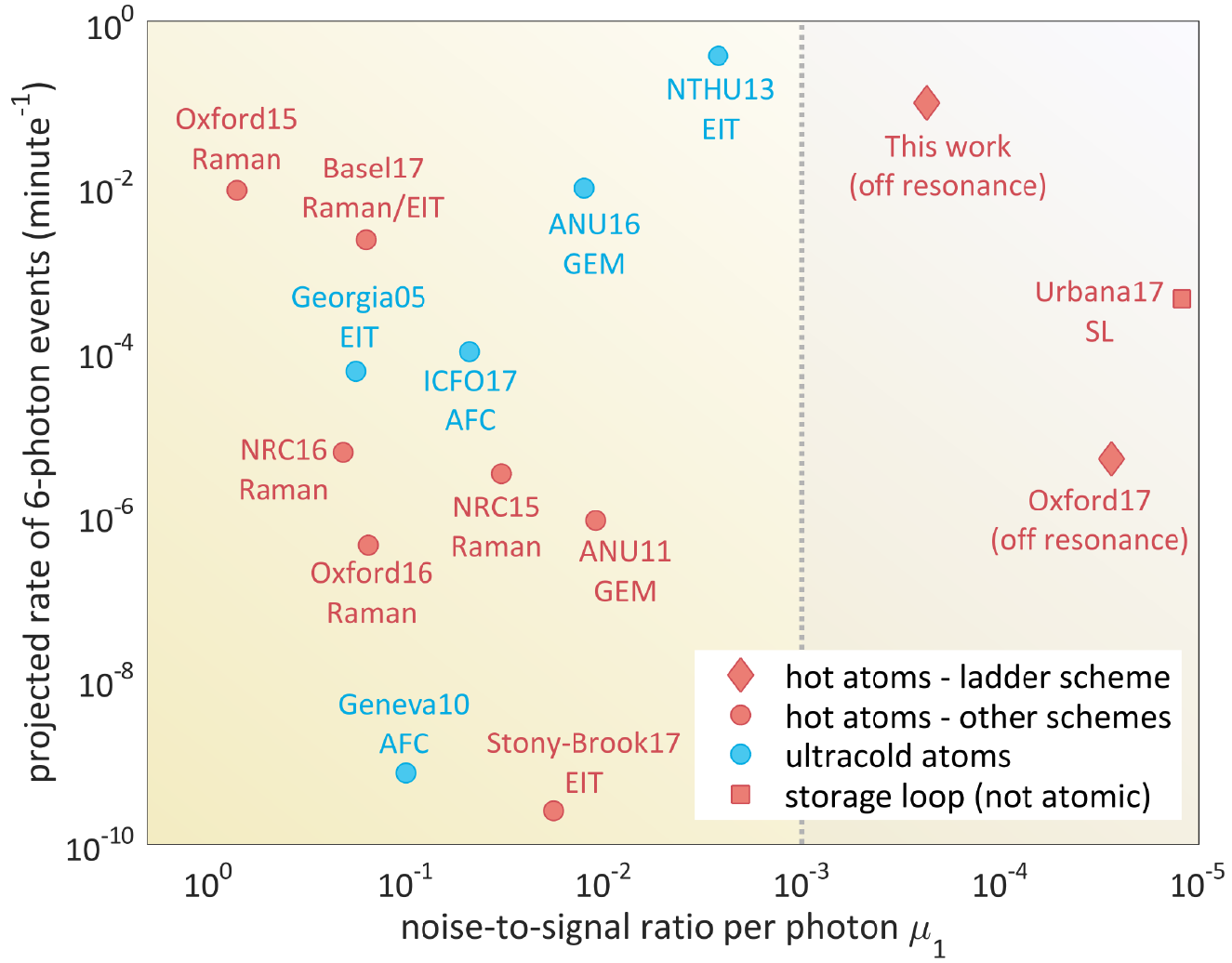}
\caption{\label{fig4} Projected performance of reported quantum memories for synchronizing 6 heralded single photon sources: high production rate and low noise (top-right quarter) are desired. The vertical dotted line marks the value $10^{-3}$ used as the success probability of the sources (thus indicating their intrinsic noise-to-signal ratio). The labels present the group, year, and memory protocol [gradient-echo memory (GEM), full atomic frequency-comb (AFC), far-detuned Raman storage (Raman), storage loop (SL), and EIT storage (EIT)]. The calculation takes the source repetition rate as the minimum between $\tp^{-1}$ and 50~GHz, the latter estimating the current limit of the required feed-forward electronics. Source data, references, and additional details are provided in Sec.~\ref{SM_comparison}.}
\end{figure}

As a benchmark, we examine a test case of synchronizing 6 probabilistic single-photon sources having initial success probability of $10^{-3}$, following Ref.~\cite{Nunn13}. Fig.~\ref{fig4} summarizes the projected performance of reported memory protocols, demonstrating the FLAME advantage. 

In conclusion, we have demonstrated FLAME in rubidium, both on and off resonance, and shown that very low noise levels, high external efficiency, and high fractional delay, are all simultaneously achievable at ambient temperature. There is much room for improvement of rubidium FLAME beyond our initial demonstration. Specifically, we estimate that the control power and optical depths required for storage of 200~ps pulses with 50\% external efficiency are within current experimental reach ({see Materials and Methods}). These parameters could enable the synchronization of, {\it e.g.}, 10 probabilistic single-photon sources in less than one second, paving the way to quantum information processing with large quantum states of light. Furthermore, by coupling it to strongly-interacting Rydberg states \cite{PfauLowRydbergPolaritonsPRA2016}, FLAME can potentially be utilized for building deterministic quantum gates or sources.

\newpage
{\section*{Materials and Metohds}
\subsection*{Experimental Design}
A detailed scheme of the experimental setup is presented in Fig.~\ref{figS1}. The core of the setup comprises a 780 nm distributed Bragg reflector (DBR) diode laser, serving as the signal beam, and a 776 nm external cavity diode laser (ECDL) amplified by a tapered amplifier (TA), serving as the control beam. The signal laser is offset-locked to a master ECDL using a fast beat-note detector, while the master laser is polarization-locked to a reference $^{85}$Rb cell. The control laser and the master laser, modulated by a fiber electro-optic phase modulator (EOPM), counter-propagate through a $^{87}$Rb reference cell, and the control laser is locked to the two-photon absorption or transparency feature. In this configuration, the frequencies of the signal and control can be independently tuned, while being locked to the master laser, whereas their sum -- and thus the two-photon detuning -- is insensitive to slow frequency drifts of the master laser.

The signal field is amplitude modulated in time by two fiber electro-optic amplitude-modulators (EOAMs) to carve a Gaussian pulse of 1.7 ns FWHM (for the off-resonance measurements) or 1.85 ns FWHM (for the on-resonance measurements), with a combined extinction ratio of 1:3000. The control field is amplitude modulated by two Pockels cells (PCs), followed by an acousto-optic modulator (AOM), to create two pulses of 2.5 ns FWHM (for the off-resonance measurements) or 10 ns FWHM (for the on-resonance measurements), with a combined extinction ratio of 1:10000 outside the two pulses, and of between 1:200 and 1:10000 between the pulses, depending on the storage time (due to the 20 ns switching time of the AOM). The repetition rate of the experiment is set by that of the PCs to 100~kHz.

After the modulators, the control beam is passed through a tilted 780 nm bandpass filter  (Semrock LL01-780-12.5) functioning as a 776 nm bandpass filter, filtering out other frequencies that might be produced in the TA. The signal and control pulses are passed through single mode fibers (SMF), aligned with each other in a counter-propagating geometry, and overlapped at the center the vapor cell. The signal (control) beam is focused down to $2w_0=170~\mu$m ($410~\mu$m) waist diameter ($1/e^{2}$). Both beams are $\sigma^+$ polarized. The control peak power used in the measurements shown in Fig.~\ref{fig2} for off (on) resonance is 200 mW (580 mW), with a corresponding peak Rabi frequency of $2\pi\times410$~MHz ($2\pi\times700$~MHz). 

The 10-mm-long $^{87}$Rb vapor cell is anti-reflection coated. It is heated to 72$^\circ$C at its coldest spot and 100$^\circ$C at the hottest spot using two electrical current heaters, to set a Rb density of $6.5\times10^{11}$~cm$^{-3}$ and an optical depth OD$\approx9$. We obtain OD$\approx15$ with continuous optical pumping and OD$\approx13$ (during the storage) when the optical pumping is turned off for storage.

After the cell, the signal beam is passed through a polarizing beam-splitter and two 780 nm bandpass filters (at normal incidence) to filter out any residual 776 nm and 795 nm components. It is then coupled to a SMF acting as a spatial filter, removing most of the spatially incoherent fluorescence emitted from the cell at 780 nm. The SMF is coupled either to a fast linear avalanche photo detector (APD) with bandwidth of 1~GHz, or to a single photon counting module (SPCM) connected to a time tagger with time bins of 100~ps.

Two optical pumping beams, a `pump' and a `repump', are introduced into the system for a duration of 8.2~$\mu s$ out of the 10~$\mu s$ period of the experimental cycle. These beams are both at 795 nm, $\sigma^+$ polarized, and separated one from the other by 6.8 GHz, such that the pump (repump) is resonant with the $F=2 \rightarrow F'=2$ ($F=1 \rightarrow F'=2$) transitions of the rubidium D1 line. The pump beam originates from a temperature- and current-stabilized DBR diode laser amplified by a TA, and the repump beam originates from a continuous-wave Ti:Sapph laser locked to a reference cavity. The pump beam power at the vapor cell is 100~mW, and that of the repump beam is 250~mW. 
The pump beam is 1.2 mm wide, and the repump beam is 1.6 mm wide. They are directed at small angles with respect to the control beam. To minimize the noise due to collisional fluorescence from atoms excited by the optical pumping (see Sec.~\ref{SM_noise} below), the experiment is performed 1.2~$\mu$s after the optical-pumping beams have been switched off by AOMs. A $\sim$1~G magnetic field applied along the optical axis and a two-layer $\mu$-metal magnetic shield protect the spin polarization from dephasing and thus from depumping due to ambient magnetic fields.

For the storage experiments presented in Fig.~\ref{fig2} of the main text and in Fig.~\ref{fig_pumping}, each data point was collected over 1 million experimental cycles using the SPCM. For the noise measurements presented in Fig.~\ref{fig3}B of the main text, each data point was collected over 5 million cycles using the SPCM. The measurement of efficiency versus control power in Fig.~\ref{fig3}A was collected using the fast APD and with weak laser pulses.

\begin{figure*}[tb]
\centering
\includegraphics[width=\textwidth]{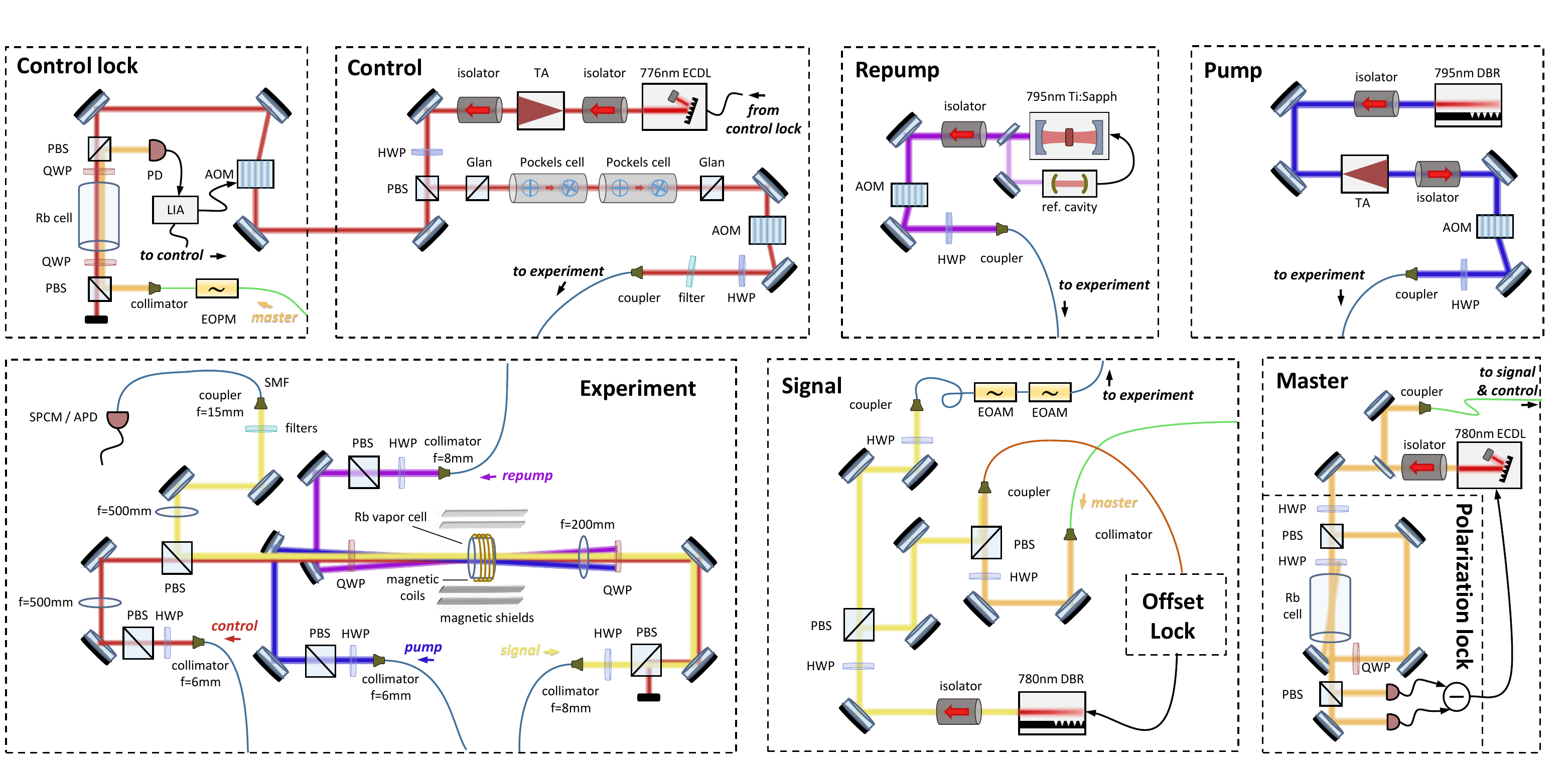}
\caption{\label{figS1} Experimental setup. QWP - quarter wave plate. HWP - half wave plate. PD - photo-detector. APD - avalanche photo-diode. SPCM - single-photon counting module.  PBS - polarizing beam splitter. Glan - Glan-laser polarizer. LIA - lock-in amplifier. TA - tapered amplifier. AOM - acousto-optic modulator. EOAM - electro-optic amplitude-modulator. EOPM - electro-optic phase-modulator. DBR - distributed-Bragg reflector laser. ECDL - external-cavity diode-laser.}
\end{figure*}

\subsection*{Analysis of memory efficiency}\label{SM_eff}
The external efficiency is the product of the internal efficiency and the setup transmission. The internal efficiency is calculated as the ratio of the area of the retrieved pulse to that of the signal pulse transmitted through the system without the control beams (at a detuning of 1.15~GHz, where its absorption is $<10^{-3}$ ). The area is summed over an integration window of 5.5~ns (8~ns) for off (on) resonance operation, containing the entire retrieved pulse.

The setup transmission is measured as the ratio of the input signal intensity just before the vapor cell, and the signal intensity reaching the detector after passing the entire system, including the bandpass filters and the SMF. The total setup transmission in our system is measured to be $T_\mathrm{setup}=78.0(5)\%$, with the filters transmitting $97\%$, the SMF coupling and transmission being $90.0(4)\%$, and the rest of the system (vapor cell and other optics) transmitting $89.0(4)\%$.

The decay of coherence during storage stems from both homogeneous and inhomogeneous processes, represented respectively by the decay times $\tau_\sigma$ and $\tau_\gamma$. The envelope decay is described by $\eff_0e^{-(t-t_0)/\tau_\gamma}e^{-(t-t_0)^2/(2\tau_\sigma^2)}$, where $\eff_0=\eff(t_0)$ is the short-time external storage efficiency. We identify the $1/e$ memory lifetime $\ts$ and a complementary parameter $\taub$ by the relations $\tau_\gamma=\ts \taub / (\taub - \ts ) $ and $\tau_\sigma=\sqrt{\ts \taub /2 }$.
The solid lines in Fig.~\ref{fig2}C,F of the main text are fits to the data using the model
\begin{equation}\label{eff_eq2}
\begin{split}
\eff(t)=&\eff_0 e^{- \left[ \frac{1}{\ts \taub} \left( t-t_0 - \ts \right) \left( t- t_0 + \taub \right) +1 \right] } \\
& \times \left| \frac{ 1+Ae^{-i\omega_{43}(t-t_0) }+Be^{-i\omega_{42}(t-t_0) } } {1+A+B} \right|^2
\end{split}
\end{equation}
where the hyperfine frequency differences within the 5D$_{5/2}$ level are taken to be $\omega_{43}=2\pi\times28.82$~MHz, and $\omega_{42}=2\pi\times51.77$~MHz~\cite{RbData}. The 5D$_{5/2}$ $F=1$ level is neglected in this model.

As the stored pulse duration is much shorter than the inverse of the frequency differences within the 5D$_{5/2}$ level, the stored coherences of different hyperfine states are all in-phase at the time of storage, $t_0$. We are thus able to find $t_0$, which might be offset from the peak of the signal reference due to modified group velocity on and off resonance, by fitting Eq.~(\ref{eff_eq2}) to the measured oscillations. The obtained fit parameters are summarized in Table~\ref{TableFitParams}.
The lifetime is governed by the inhomogeneous (motional) dephasing $\tau_\sigma=\sqrt{\ts \taub /2 }$, and we can extract from the fit $\tau_\sigma=65(4)$ ns for the off-resonance case and $\tau_\sigma=117(7)$ for the on-resonance case. The much longer homogeneous decay time cannot be faithfully determined from the data.

\begin{table*}[tb]
\def\arraystretch{1.4}%
\centering
\caption{Memory parameters extracted from the measurements using the fit function of Eq.~(\ref{eff_eq2}). Uncertainties in parentheses are 1$\sigma$ standard deviation.
}\label{TableFitParams}

\begin{tabular}{|c||c|c|c|c|c|c|c|}
\hline
& $\ts$ (ns) & $\eff_0$ (\%) & $\taub$ (ns) & $t_0$ (ns) & $A$ & $B$ \\
\hline
\hline
Off-resonance  & 86(2) & 25.1(8) &  101(12)   &  -1.0(3) & 0.160(9) & 0.006(9) \\
\hline
On-resonance & 82(1) &  17.1(3) &  337(43)  & 9.2(6) & 0.032(4) & 0.007(4) \\
\hline   
\end{tabular}

\end{table*}

The main contribution to the storage decay rate is due to ballistic thermal motion of the hot atoms. This includes both the longitudinal residual Doppler broadening $\Delta k v_T =1.22~(2\pi)~\mathrm{MHz}$ and the transversal `time-of-flight' rate of the atoms leaving the beam $v_T /w_0=0.34~(2\pi)\mathrm{MHz}$, with $w_0$ the signal beam waist radius. The corresponding inhomogeneous decay time is 102 ns. Summing these rates with the homogeneous `natural' coherence decay rate $0.33~(2\pi)\mathrm{MHz}$ of the $5D_{5/2}$ level, the storage lifetime is estimated as 84 ns, consistent with the above results. We believe that the lower inhomogeneous lifetime $\tau_\sigma$ for the off-resonant case is related to its high bandwidth, addressing most atoms including those with high thermal velocity. Understanding the details of this dephasing mechanism in the context of the differences between the on- and off-resonance cases will be part of future work.

In accordance with the above values, the short-time internal efficiency for $\Delta=1.15~$GHz is $\eta^{\mathrm{int}}_0=32(1)\%$, 
and for $\Delta=0$, it is $\eta^{\mathrm{int}}_0=22.0(4)\%$.
For the off-resonance regime, we can roughly estimate the expected storage efficiency using the formalism developed in Ref.~\cite{NunnPhD2008}. We use $\gamma\times\mathrm{\ODs}=5~(2\pi)~$GHz [$\gamma=6~(2\pi)~$MHz the spectral width of the intermediate level and $\ODs$ the optical depth had the atoms were stationary] and $\Omega/\Delta=0.36$ to calculate for a pulse width $\tp=1.7~$ns the so-called coupling parameter $\mathcal{C}=(\Omega/\Delta)\sqrt{\tp\gamma\ODs}/4=0.66$. The resulting (internal) efficiency is $\eta^{\mathrm{int}}_0(\mathcal{C})\approx 16\%$ \cite{NunnPhD2008}, which is lower but qualitatively agreeing with the measured value. The same formalism implies that for control intensity and optical depth an order of magnitude larger than those available in our current setup 
, storage of 200-ps pulses with $50\%$ efficiency ($\mathcal{C}\approx 1$) is possible. {This estimation assumes: a control power of 6 W (30-times larger than the 200 mW used for the above calculation, so that $\Omega\rightarrow \sqrt{30}\Omega$) readily available with current laser technology such as continuous Ti:Sapphire lasers; larger detuning ($\Delta \rightarrow 3\Delta$) for avoiding resonant absorption of the broadband pulses; and a density 10-times larger ($\ODs \rightarrow 10\times \ODs$), readily achievable by heating the cell to $105^\circ$C.}}
\section*{Acknowledgments:}
 We thank J.~Nunn for stimulating discussions. We acknowledge financial support by the Israel Science Foundation and ICORE, the European Research Council starting investigator grant Q-PHOTONICS 678674, the Minerva Foundation, the Sir Charles Clore research prize, and the Laboratory in Memory of Leon and Blacky Broder. All authors contributed to the experimental design, construction, data collection, and analysis of this experiment. R.F. claims responsibility for figures 1-3,S1,S2; E.P. and O.F. produced Fig. 4 and its data; O.L. produced Fig. 5. The authors wrote the manuscript together. All authors declare that they have no competing interests. All data needed to evaluate the conclusions in the paper are present in the paper. Additional data related to this paper may be requested from the authors.
\bibliography{RbLadder}
\clearpage

\renewcommand\theequation{S\arabic{equation}}
\setcounter{equation}{0}  
\renewcommand\thefigure{S\arabic{figure}}
\setcounter{figure}{0}  
\renewcommand\thetable{S\arabic{table}}
\setcounter{table}{0}  
\renewcommand\thesubsection{S\arabic{subsection}}
\setcounter{subsection}{0}  
\renewcommand\thesubsubsection{\thesubsection.\arabic{subsubsection}}

\newpage 

\section*{Supplementary material}

\label{SM_exp}
\subsection{Atomic level scheme}
\begin{figure}[b]
\centering
\includegraphics[width=\columnwidth]{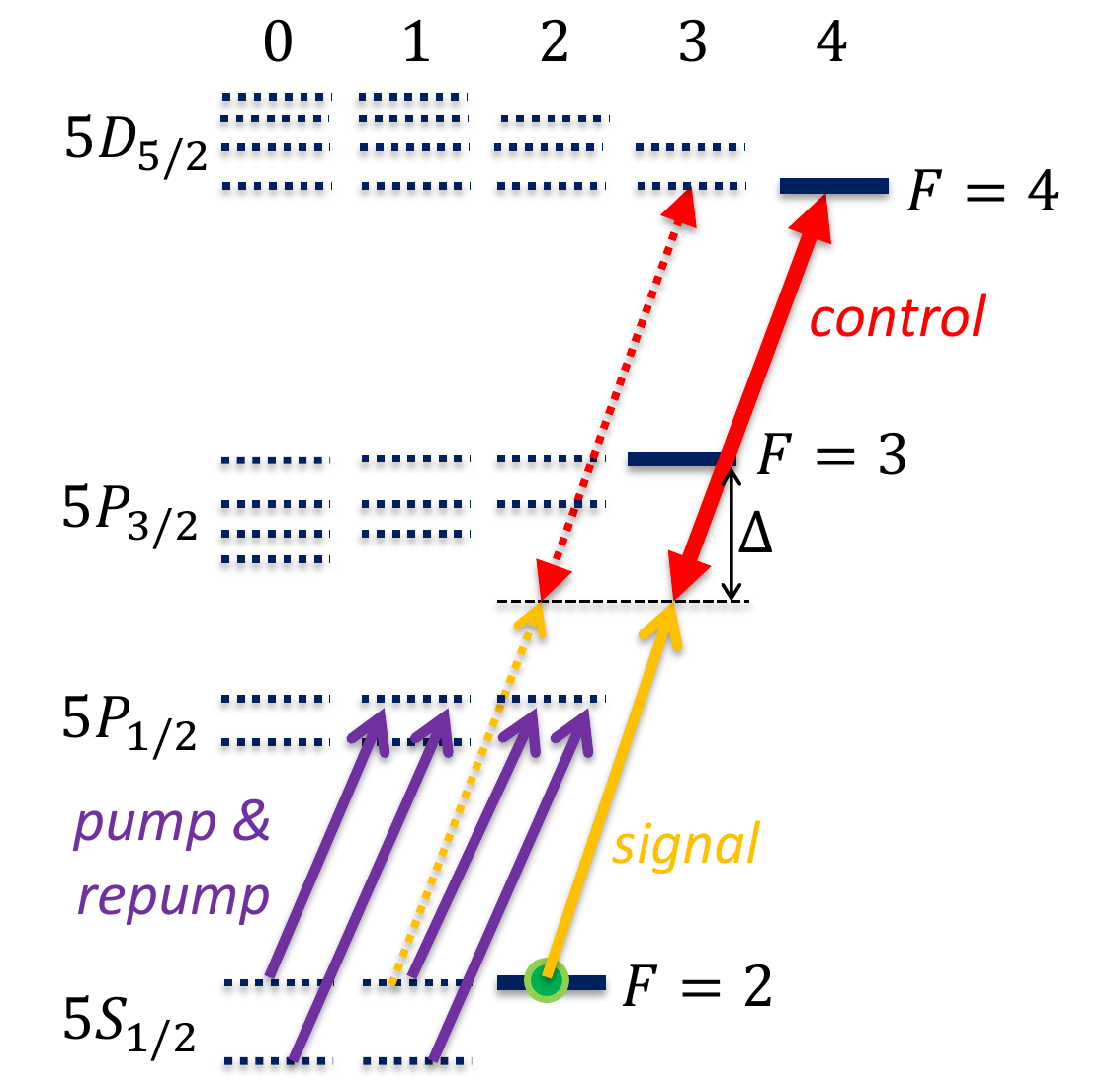}
\caption{\label{fig_scheme} Rubidium level scheme. Colored arrows represent the optical fields.}
\end{figure}

The energy levels and sub-levels involved in the experiment are shown in Fig.~\ref{fig_scheme}. The $\sigma^+$ polarized pump and repump (purple arrows) excite all 5S$_{1/2}$ states except for the maximally-polarized state $|F=2;m_F=2\rangle$, moving population to that state. If all the atoms are in the $|F=2;m_F=2\rangle$ state, $\sigma^+$ polarized signal (yellow) and control (red) can excite the system only to the 5P$_{3/2}|F=3;m_F=3\rangle$ and 5D$_{5/2}|F=4;m_F=4\rangle$ states (solid arrows). If the optical pumping is not perfect,  other ground states are populated, and other excited states can be reached (dotted arrows). 

\begin{figure}[tb]
\centering
\includegraphics[width=\columnwidth]{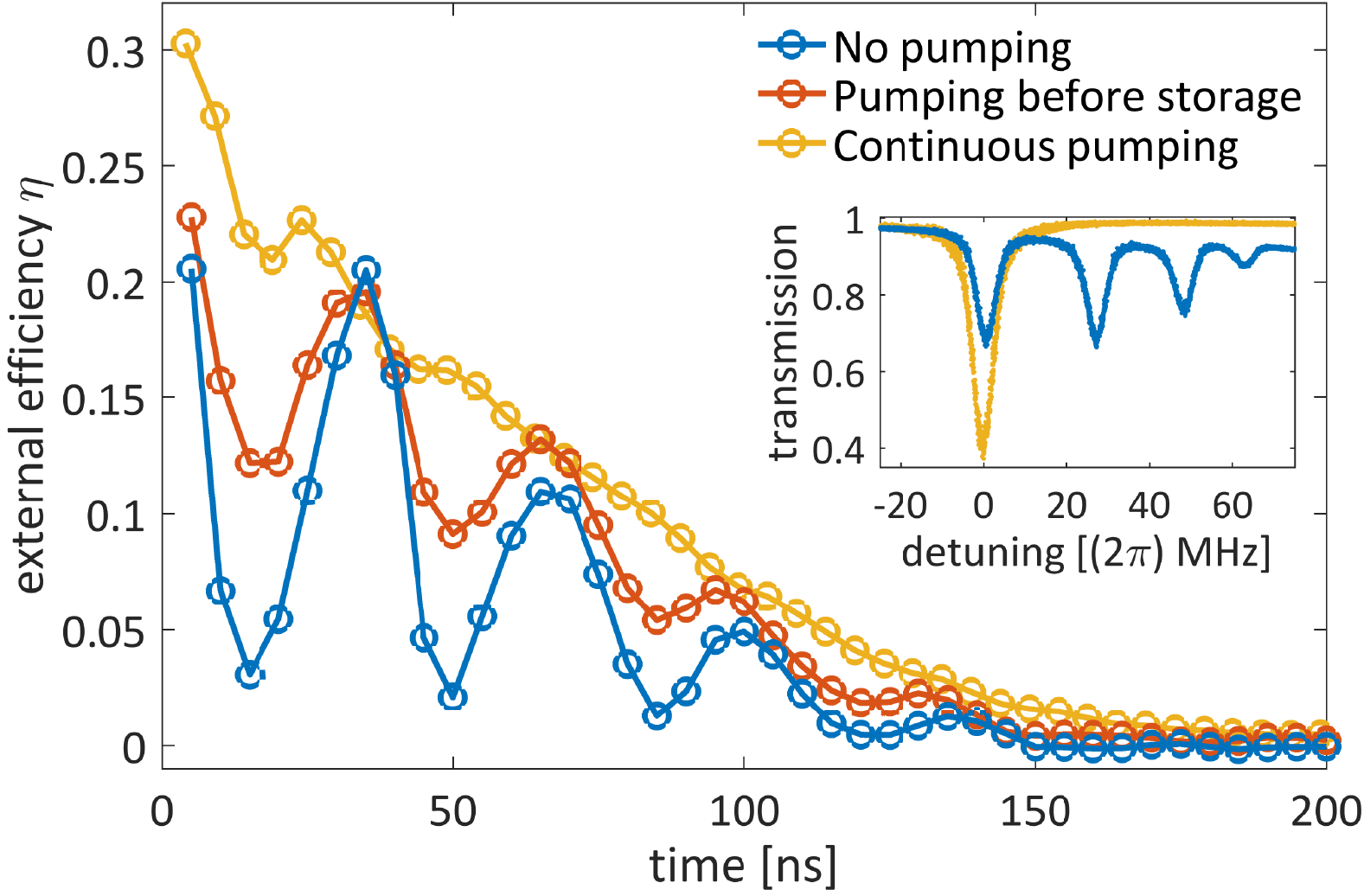}
\caption{\label{fig_pumping}External efficiency as a function of storage time $t$, with the optical pumping continuously on (yellow) or switched off before storage (red), or with no pumping at all (blue). Inset: Measured two-photon absorption spectrum with (yellow) and without (blue) optical pumping. }
\end{figure}

Figure \ref{fig_pumping} shows the effect of optical pumping. Without optical pumping, the spectrum comprises 4 absorption dips and the storage efficiency displays beats with high contrast. When optical pumping is turned on, the spectrum consists of a single dip, and the beats vanish. In the experiment, switching the pumping beams off before storage minimizes the noise, but residual beating appears in the efficiency.

\subsection{Noise sources}\label{SM_noise}
As shown in Fig.~3B of the main text, the noise comprises a component independent of control power and a smaller component which increases linearly with control power. This suggests that there are two main noise sources in our experiment. We attribute the power-dependent component to control leakage and the power-independent component to collisional fluorescence due to optical pumping.
\paragraph{Control leakage.} As described earlier, the control field is initially filtered from residual 780 nm light, originating from amplified spontaneous emission of the TA, using a tilted bandpass filter. Before the SMF going to the detector, two bandpass filters are placed to allow only 780 nm to pass and remove any stray 776 nm light; each filter attenuates this noise by 7 orders of magnitude. In addition, due to the counter-propagation geometry, only reflections of the control from optical surfaces (and the beam-dump) are in the direction of the detector. Still some small fraction of the control light reaches the detector and gives rise to noise that is linearly proportional to the control power. 

\paragraph{Collisional fluorescence due to optical pumping.} During optical pumping, a considerable fraction of the atoms populate the 5P$_{1/2}$ level. Collisions may transfer population into the 5P$_{3/2}$ level, from which 780 nm photons are spontaneously emitted~\cite{Rotondora98_Collisional_transfer} and counted as noise. 
We have verified that this fluorescence does not depend on control power. We observed a significant reduction of it at lower atomic densities, at which collisions are less frequent. However, we have also observed that the time scale over which the fluorescence decays is much longer than the $26$~ns radiative lifetime of the 5P$_{3/2}$, which calls for a more elaborate study of this process in the future.  

To reduce the noise counts due to this process at our working temperature, we use two means. First, we perform the storage 1.2~$\mu$s after the pump beams are switched off. This delay suppresses the noise by more than two orders of magnitude. Second, we use a SMF as a spatial filter, accepting only a small fraction of the isotropic fluorescence emission. We estimate this fraction by $\zeta\approx(NA)^2/4\approx3\times10^{-6}$, where the collection numerical aperture of the SMF is NA$=0.11$, divided by the demagnification factor of about 30. Overall, we suppress the collisional fluorescence noise by about 8 orders of magnitude, measuring a noise component (independent of control power) of $4.6\times10^{-5}$  photons ($8.4\times10^{-6}$ photons per ns) in the off-resonance case and $8.9\times10^{-5}$ photons ($11\times10^{-6}$ photons per ns) in the on-resonance case.

\subsection{Source data and performance analysis of different quantum-optical memories} \label{SM_comparison}
Tables \ref{Table1_for_fig4} and \ref{Table2_for_fig4} detail the parameters of the quantum-optical memories presented in Fig.~\ref{fig4} of the main text. For each memory, the source reference is given in parenthesis, and all values are annotated to indicate the source location: $MT$ - main text; $MS$ - methods section; $SM$ - supplementary material; $EF$ - extracted from a figure (figure number is stated); $C$ - calculated using the  available data (the formulas used for these calculation are stated in the table caption); $NG$ - not given, the value is an estimation for the upper limit.

Note that devices that cannot be used as general-purpose quantum memories, either because they require a pre-programmable storage time \cite{AFC1,Tittel16} or because they utilize an internal source rather than storing incoming photons~\cite{DLCZ1,Furusawa16_cavity_source,XianMin_DLCZ}, are not included in this overview.

The formula for the 6-photon rate is taken from Ref.~\cite{Nunn13} and is given by,
\begin{equation}\label{rN}
r_N=\tau_c^{-1}q^N\left(1+\frac{(1-R)(1-q)\eff_0}{b+(R+q-2Rq)(1-b)}\right)^N,
\end{equation}
where $\tau_c$ is the basic clock cycle, $N=6$ is the number of synchronized sources, \mbox{$q=0.001$} is the pair emission probability of each source, $b=1-e^{-\frac{1}{f}}$, where $f$ is the fractional delay, is the memory loss probability per clock cycle, and $\eff_0$ is the short-time external efficiency. \mbox{$R=Y^N$}, where $Y$ is the positive real root of
\mbox{$(1-2q)Y^N+q^2Y^{N-1}+qY-q=0$}. 
Thus $R=0.0024$ for $N=6$ and $q=0.001$.

\begin{turnpage}
\begin{table}[tb]
\def\arraystretch{1.3}%
\centering
\caption{Memory parameters used for compiling Fig.~\ref{fig4} of the main text. $\tp$ - signal pulse duration (FWHM). $\tau_\mathrm{c}$ - minimal clock cycle. This is the higher of $\tp$ and 20~ps, the latter being an estimation of the minimal response time of synchronization electronics. $\ts$ - 1/$e$ storage time. $f'=\ts/\tau_\mathrm{c}$ - the fractional delay, in terms of clock cycles. $f'_\mathrm{e}=f'\eff_0$ - the effective fractional delay in terms of clock cycles, where $\eff_0$ is the short-time external efficiency (see Table~\ref{Table2_for_fig4}).\vspace{2mm}}\label{Table1_for_fig4}
  \begin{tabular}{|c||c|c|c|c|c|}
   \hline
  Reference  & $\tp$ (sec) & $\tau_\mathrm{c}$ (sec) & $\ts$ (sec) & $f'$ & $f'_\mathrm{e}$\\
  \hline
  \hline
  Oxford 15 \cite{Michelberger15}& 3.6$\times10^{-10}$ $^{MT}$ & 3.6$\times10^{-10}$ $^{C}$& 1.5$\times10^{-6}$ $^{MT}$& 4167 $^{C}$ & 77 $^{C}$\\
  \hline
  Oxford 16 \cite{Saunders16} & 3.2$\times10^{-10}$ $^{MT}$ & 3.2$\times10^{-10}$ $^{C}$& 9.5$\times10^{-8}$ $^{MT}$& 297 $^{C}$ & 0.29 $^{C}$\\
  \hline
  NRC 15 \cite{Diamond_phonon_memory2}    & 2.6$\times10^{-13}$ $^{MT}$& 2$\times10^{-11}$ $^{C}$& 3.5$\times10^{-12}$ $^{MT}$& 0.175 $^{C}$ & 0.0016 $^{C}$\\
  \hline
  NRC 16 \cite{SussmanHydrogen16}  & 1.75$\times10^{-13}$ $^{MT}$& 2$\times10^{-11}$ $^{C}$& 8.5$\times10^{-11}$  $^{MT}$& 4.25 $^{C}$ & 0.1 $^{C}$\\
  \hline
  Basel 17 \cite{TreutleinRamanEIT17}  & 6.7$\times10^{-10}$ $^{MT}$& 6.7$\times10^{-10}$ $^{C}$& 6.8$\times10^{-8}$  $^{MT}$& 101 $^{C}$ & 5.8 $^{C}$\\
  \hline
  Georgia 05 \cite{KuzmichEIT05}& 1.6$\times10^{-7}$ $^{MS}$& 1.6$\times10^{-7}$ $^{C}$& 1.1$\times10^{-5}$  $^{EF 5}$& 69 $^{C}$ & 7.7 $^{C}$\\
  \hline
  NTHU 13 \cite{YuEIT13}   & 8.8$\times10^{-7}$ $^{MT}$& 8.8$\times10^{-7}$ $^{C}$& 9.2$\times10^{-5}$  $^{MT}$ $^{(a)}$& 105 $^{C}$ & 56 $^{C}$\\
  \hline
  Stony-Brook 17 \cite{StonyBrook17}   & 5$\times10^{-7}$ $^{MT}$& 5$\times10^{-7}$ $^{C}$& 1.4$\times10^{-5}$  $^{MT}$ & 28 $^{C}$ & 0.14 $^{C}$\\
  \hline
  ANU 11 \cite{BuchlerGem11_1,BuchlerGem11_2} & 3$\times10^{-6}$ $^{MT}$& 3$\times10^{-6}$ $^{C}$& 2.2$\times10^{-5}$  $^{MT}$& 7.3 $^{C}$ & 4.7 $^{C}$\\
  \hline
  ANU 16 \cite{BuchlerGem16} & 6.66$\times10^{-6}$ $^{MT}$& 6.66$\times10^{-6}$ $^{C}$& 3.3$\times10^{-4}$ $^{MT}$ $^{(b)}$ & 50 $^{C}$ & 36 $^{C}$\\
  \hline
  Geneva 10 \cite{GisinsAFC10} & 4.5$\times10^{-7}$ $^{MT}$& 4.5$\times10^{-7}$ $^{C}$& 1.6$\times10^{-6}$ $^{EF 4}$ & 36 $^{C}$ & 0.36 $^{C}$\\
  \hline
  ICFO 17 \cite{DeRidsAFC17}   & 6.3$\times10^{-8}$ $^{MT}$& 6.3$\times10^{-8}$ $^{C}$& 1.8$\times10^{-5}$ $^{EF 4a}$ & 286 $^{C}$ &  11.4 $^{C}$\\
  \hline
  Urbana 17 \cite{Kwiat17}   & 5$\times10^{-12}$ $^{MT}$& 1$\times10^{-8}$ $^{MT}$ $^{(c)}$& 8.3$\times10^{-7}$ $^{SM}$ & 83 $^{MS}$ &  6.9 $^{C}$\\
  \hline
  Oxford 17 \cite{Kris17} & 4.4$\times10^{-10}$ $^{MT}$& 4.4$\times10^{-10}$ $^{C}$& 5.4$\times10^{-9}$ $^{MT}$ & 12.3 $^{C}$ &  0.8 $^{C}$\\
  \hline
  This work $^{(d)}$ & 1.7$\times10^{-9}$ & 1.7$\times10^{-9}$ & 8.6$\times10^{-8}$  & 50.6 & 12.6\\
  \hline
  This work $^{(e)}$ & 1.85$\times10^{-9}$ & 1.85$\times10^{-9}$ & 8.2$\times10^{-8}$  & 44.3 & 7.6\\
  \hline
  \end{tabular}
  
  \small{ \vspace{2mm}
\begin{minipage}{0.7\paperwidth}
  $^{(a)}$ For backwards retrieval.\\
  $^{(b)}$ For 0.2$^{\circ}$ between signal and control - when control leakage is sufficiently suppressed.\\
  $^{(c)}$ Determined by the storage loop round-trip time, limited by the switching time.\\ 
  $^{(d)}$ Off-resonance storage.\\
  $^{(e)}$ On-resonance storage.\\
  \end{minipage}
  }
\end{table}
\begin{table}[tb]
\def\arraystretch{1.3}%
\centering
\caption{Memory parameters used for compiling Fig.~\ref{fig4} of the main text (cont.).  $\eta^{\mathrm{int}}_0$ - internal efficiency for zero storage time. $T_\mathrm{setup}$ - setup transmission. $\eff_0=\eta^{\mathrm{int}}_0 T_\mathrm{setup}$ - external efficiency for zero storage time. $\noise$ - number of noise photons arriving at the detector per retrieval attempt. 
$\nsr=\noise/\eff_0$
- number of input photons for which the output signal-to-noise ratio at zero storage time is 1. $r_{6}$ - calculated 6 photon rate from Eq.~(\ref{rN}).\vspace{2mm}}\label{Table2_for_fig4}
  \begin{tabular}{|c||c|c|c|c|c|c|}
   \hline
  Reference & $\eta^{\mathrm{int}}_0$ (\%)& $T_\mathrm{setup}$ (\%) & $\eff_0$ (\%) & $\noise$ & $\nsr$ & $r_{6}$ (min$^{-1}$) \\
  \hline
  \hline
  Oxford 15 \cite{Michelberger15}& 21 $^{MT}$  & 8.8 $^{SM}$ & 1.85 $^{C}$ & 0.013 $^{MT}$ $^{(a)}$& 0.71 $^{C}$ & 0.009 $^{C}$\\
  \hline
  Oxford 16 \cite{Saunders16}& 9.7 $^{C}$ $^{(b)}$ & 1 $^{MT}$ & 0.097 $^{C}$ & 1.5$\times10^{-4}$ $^{MT}$ $^{(a)}$& 0.16 $^{MT}$ & 4$\times10^{-7}$ $^{C}$\\
  \hline
  NRC 15 \cite{Diamond_phonon_memory2}& 0.9 $^{MT}$& 100 $^{NG}$ & 0.9 $^{C}$& 3$\times10^{-4}$ $^{EF 2}$ & 0.033 $^{C}$& 3.2$\times10^{-6}$ $^{C}$\\
  \hline
  NRC 16 \cite{SussmanHydrogen16}& 10.1 $^{MT}$&22.6 $^{MT}$& 2.4 $^{MT}$& 0.0045 $^{MT}$& 0.21 $^{EF 3}$ & 5.7$\times10^{-6}$ $^{C}$\\
  \hline
  Basel 17 \cite{TreutleinRamanEIT17}& 17 $^{MT}$ & 34 $^{MT}$ & 5.8 $^{C}$ & 0.0092 $^{MT}$ $^{(a)}$& 0.16 $^{C}$& 0.0022 $^{C}$\\
  \hline
  Georgia 05 \cite{KuzmichEIT05}& 41.5 $^{EF 3}$ $^{(b)}$ &27 $^{SM}$ &11.2 $^{C}$ & 0.021 $^{SM}$ $^{(a)}$& 0.19  $^{C}$& 5.6$\times10^{-5}$ $^{C}$\\
  \hline
  NTHU 13 \cite{YuEIT13}& 78 $^{MT}$ $^{(c)}$& 68 $^{MT}$ $^{(c)}$& 53 $^{C}$& 0.0014 $^{MT}$ $^{(a)}$ $^{(d)}$ & 0.0026 $^{C}$ & 0.38 $^{C}$\\
  \hline
  Stony-Brook 17 \cite{StonyBrook17}& 11 $^{MT}$ & 4.5 $^{MT}$ & 0.5 $^{C}$& 9$\times$10$^{-5}$ $^{EF3,5}$  & 0.018 $^{C}$ & 2.5$\times$10$^{-10}$ $^{C}$\\
  \hline
  ANU 11 \cite{BuchlerGem11_1,BuchlerGem11_2}& 91 $^{EF 4a}$ $^{(e)}$& 70 $^{SM}$ $^{(f)}$ & 64 $^{C}$ & 0.0064 $^{EF 3}$ $^{(a)}$ $^{(e)}$ & 0.011 $^{C}$ & 8.4$\times10^{-7}$ $^{C}$\\
  \hline
  ANU 16 \cite{BuchlerGem16}& 80 $^{EF 4a}$ & 90 $^{SM}$ & 72 $^{C}$ & 0.009 $^{EF 5a}$ $^{(a)}$& 0.0125 $^{C}$& 0.0094 $^{C}$\\
  \hline
  Geneva 10 \cite{GisinsAFC10}& 1 $^{MT}$ & 100 $^{NG}$ & 1 $^{C}$ & 0.001 $^{EF 4}$ & 0.1 $^{C}$ & 7.1$\times10^{-10}$ $^{C}$\\
  \hline
  ICFO 17 \cite{DeRidsAFC17} & 4 $^{MT}$ & 100 $^{NG}$ & 4 $^{C}$ & 0.0019 $^{MT}$ & 0.048 $^{C}$ & 9.7$\times10^{-5}$ $^{C}$\\
  \hline
  Urbana 17 \cite{Kwiat17} & 98.8 $^{MT}$ & 8.3 $^{SM}$& 8.3 $^{C}$& 1$\times10^{-6}$ $^{(g)}$& 1.2$\times10^{-5}$ $^{C}$ & 0.0004 $^{C}$\\
  \hline
  Oxford 17 \cite{Kris17} & 22.2 $^{MT}$ $^{(b)}$ & 30 $^{SM}$ & 6.6 $^{C}$ & 1.8$\times10^{-6}$ $^{MT}$ $^{(a)}$ $^{(g)}$& 2.7$\times10^{-5}$ $^{C}$& 4.8$\times10^{-6}$ $^{C}$\\
  \hline
  This work $^{(h)}$& 32.2 & 78 & 25.1 & 5.8$\times10^{-5}$ & 2.3$\times10^{-4}$  & 0.102 $^{C}$\\
  \hline
  This work $^{(i)}$& 22 & 78 & 17.1 & 1.9$\times10^{-4}$ & 0.0011  & 0.0066 $^{C}$\\
  \hline
  \end{tabular}

  \small{ \vspace{2mm}
\begin{minipage}{0.5\paperwidth}
  $^{(a)}$ Noise right after memory multiplied by setup transmission\\
  $^{(b)}$ Extrapulated to 0 delay assuming Gaussian dependence.\\
  $^{(c)}$ For backwards retreival.\\
  $^{(d)}$ Theoretical prediction.\\
  $^{(e)}$ Ref. \cite{BuchlerGem11_2}\\
\end{minipage}
\begin{minipage}{0.3\paperwidth}
  $^{(f)}$ Ref. \cite{BuchlerGem11_1}\\
  $^{(g)}$ Limited by detector dark counts.\\
  $^{(h)}$ Off-resonance storage.\\
  $^{(i)}$ On-resonance storage.\\
\end{minipage}
  }
\end{table}
\end{turnpage}





\end{document}